\documentclass[sigconf]{acmart}

\usepackage{xspace}
\usepackage{pifont}
\usepackage{listings}
\usepackage{multirow}
\usepackage{booktabs}
\usepackage{tcolorbox}

\newcommand{\codefont}[1]{\footnotesize{\texttt{#1}}\normalsize}
\newcommand{\totalthreads}{130\xspace}
\newcommand{\totalJava}{407\xspace}
\newcommand{\totalanswer}{2,323\xspace}
\newcommand{\totalnonref}{2,137\xspace}
\newcommand{\totalref}{186\xspace}
\newcommand{\totalentry}{430\xspace}
\newcommand{\totalproj}{357\xspace}

\AtBeginDocument{%
  \providecommand\BibTeX{{%
    \normalfont B\kern-0.5em{\scshape i\kern-0.25em b}\kern-0.8em\TeX}}}

\setcopyright{acmcopyright}
\copyrightyear{2018}
\acmYear{2018}
\acmDOI{XXXXXXX.XXXXXXX}

\acmConference[Conference acronym 'XX]{Make sure to enter the correct
  conference title from your rights confirmation emai}{June 03--05,
  2018}{Woodstock, NY}
\acmPrice{15.00}
\acmISBN{978-1-4503-XXXX-X/18/06}

\begin{document}

\author{Juntong Chen}
\affiliation{
  \institution{Virginia Tech}            %% \institution is required
  \country{USA}                    %% \country is recommended
}
\email{j6xjun@gmail.com}          %% \email is recommended

\author{Yan Zhao}
\affiliation{
  \institution{Eastern Michigan University}            %% \institution is required
  \country{USA}                    %% \country is recommended
}
\email{yanzhao@vt.edu}  

\author{Na Meng}
\affiliation{
  \institution{Virginia Tech}            %% \institution is required
  \country{USA}                    %% \country is recommended
}
\email{nm8247@cs.vt.edu}          %% \email is recommended

%\title{Characterizing The Code Flow from StackOverflow To Github}
\title{How Do Developers Reuse StackOverflow Answers in Their GitHub Projects?}

\begin{abstract}

StackOverflow (SO) is a widely used question-and-answer (Q\&A) website for software developers and computer scientists. GitHub is an online development platform used for storing, tracking, and collaborating on software projects. 
Prior work relates the information mined from both platforms to link user accounts or compare developers' activities across platforms. However, not much work is done to characterize the SO answers reused by GitHub projects. 
For this paper, we did an empirical study by mining the SO answers reused by Java projects available on GitHub. 
%Because each SO discussion thread can have one or multiple answers, 
We created a hybrid approach of clone detection, keyword-based search, and manual inspection, to identify the answer(s) actually leveraged by developers.
Based on the identified answers, we further studied topics of the discussion threads, answer characteristics (e.g., scores, ages, code lengths, and text lengths), and developers' reuse practices.

We observed that most reused answers offer programs to implement specific coding tasks. Among all analyzed SO discussion threads, the reused answers often have relatively higher scores, older ages, longer code, and longer text than unused answers.  In only 9\% of scenarios (40/\totalentry), developers fully copied answer code for reuse. In the remaining scenarios, they reused partial code or created brand new code from scratch. 
Our study characterized \totalthreads SO discussion threads referred to by Java developers in \totalproj GitHub projects. Our observations can guide SO answerers to provide better answers, and shed lights on future human-centric software engineering research that creates better tools to facilitate reliable and responsible code reuse. 

\end{abstract}

%\textbf{Background:} 
%\textbf{Aims:} 
%\textbf{Method:}

%\textbf{Results:} 
%\textbf{Conclusions:} 


%\ccsdesc[100]{Software and its engineering~Software configuration management and version control systems}


\maketitle

\section{Introduction}
Each month, about 50 million people visit StackOverflow (SO) to learn, share, and build their careers. Industry estimates suggest that 20--25 million of these people are professional developers and univerity-level students~\cite{2020-developer-survey}. In February 2020, a survey with nearly 65,000 developers shows that when stuck on a coding problem, 90\% of respondents visited SO. The fact implies that SO plays an important role in shaping the art and practices of software today. 

Studies were conducted to characterize the crowdsourced knowledge available on SO, or to relate the mined knowledge from SO and from GitHub~\cite{hello-world}. For example, Nasehi et al.~\cite{Nasehi2012} manually inspected 163 unique SO posts with well-received answers; they found that the explanations accompanying examples are as important as the code examples themselves. Vasilescu et al.~\cite{Vasilescu13} identified GitHub developers active on SO, to study their activities on both platforms. They observed that GitHub committers ask fewer questions and provide more answers than others. Manes and Baysal~\cite{Manes2019} mined the SOTorrent dataset; they found 
 that on average, developers make 45 references to SO posts in their GitHub projects, with the highest number of references made in JavaScript code.

However, little is known concerning how developers leverage the crowdsourced knowledge available on SO, and what kind of SO posts got referenced by GitHub projects. 
Such knowledge can guide answerers to provide better answers, assist questioners to better compare answers, and shed light on new tools that recommend customized coding solutions to developers based on SO answers. For this paper, we did an empirical study to explore the following research questions to complement prior work:

%; active SO askers distribute their work in a less uniform way than developers that do not ask questions. 

% They also found that 79\% of the SO posts with code snippets referenced by GitHub do change over time. 
\begin{itemize}
\item[\textbf{RQ1}] \emph{What kinds of SO discussion threads are referenced by GitHub projects?} When certain discussion threads have URLs cited by GitHub projects, we aimed to characterize the discussion topics. Those topics can reflect developers' focus when seeking for coding assistance; thus, they can guide researchers, tool builders, and SO answerers to better help developers. 
%to provide better coding assistance and create higher impacts.
%They can guide researchers and tool builders to create better coding assistance tools, and enable SO answerers to better get involved into discussion threads and potentially create higher impacts. 
\item[\textbf{RQ2}] \emph{What are the characteristics of reused SO answers?} When an SO discussion thread has multiple alternative answers, we aimed to characterize the factors (e.g., scores and code lengths) that may influence developers' decisions on choosing some answers over the others. Such characteristics can guide SO answerers to provide answers with higher quality, and to earn reputation more effectively.
%\item[\textbf{RQ2}] \emph{What kinds of code snippets are copied from SO and reused by GitHub projects?} We aimed to identify the common characteristics of code answers reused by GitHub projects. Such characteristics can guide SO answerers to provide code examples with higher quality, to help developers more effectively.
\item[\textbf{RQ3}] \emph{How are SO answers reused in GitHub projects?} 
When referring to SO answers, developers may reuse some or all exemplar code, revise the code as needed, or create totally new code based on the insights. 
%When copying code from SO, developers may use the copied code as is or revise it as needed. 
By characterizing the answer-reuse practices of GitHub developers, we intended to study how SO answers help shape software products.
\end{itemize}

To investigate the research questions mentioned above, we first used a fully managed enterprise data warehouse Google BigQuery~\cite{bigquery}, to crawl for Java files in GitHub projects that reference any SO discussion thread (i.e., URL), and to download 30,000 answer posts contained by all those referenced SO discussion threads. 
We chose Java because it is widely used and we are more familiar with the language; such a familarity enables us to manually analyze SO posts and GitHub code with high confidence.
An \textbf{SO discussion thread} typically has a question post and one or more answer posts. As each post has a URL, developers may reference a discussion thread via the URL of question and/or any answer in that thread. 
For each of the 30,000 downloaded answers, we extracted exemplar code, text (i.e., code + natural-language explanation), and descriptive metadata (e.g., scores and creation timestamps). 

%As each referenced SO discussion thread may contain one or more answers reusable by GitHub projects, 
Based on the crawled data, 
we applied a novel hybrid approach to decide which answer(s) in a thread were actually used by developers. This approach has three steps. In Step 1, it uses a clone detection tool---PMD~\cite{pmd}---to find any answer that has code similar to the Java code on GitHub.
Our insight is that if developers reference an SO thread in their code, it is likely that they create certain duplicates of the answer code from that thread. Step 2: 
as PMD may be insufficient to locate all answer reuses, we also extracted the IDs of all answer posts in our dataset. We searched for Java files that explicitly mention any of these IDs, considering them as indicators for answer reuse. Step 3: we manually inspected the GitHub$\rightarrow$SO reference links revealed by Steps 1 and 2, to refine the dataset.
%to compare the answers and GitHub code and to remove false positives in established links.  
% and to determine which answers were actually reused by developers. 
%to decide which answer(s) in a thread were actually used by developers. Specifically for each referenced SO thread $T$, we extracted code snippets from all answers, and compared those snippets with the Java file referencing $T$ to locate code clones. If a pair of clones is found between the code snippet of an answer $A$ and the Java file, we consider $A$ to be actually used by developers. 

%With the referenced answers identified in this way, we further crawled the StackOverflow data dump~\cite{datadump} for their descriptive metadata such as scores, creation time, 

%With \totalref reused answers identified using the approach mentioned above, we manually analyzed the \totalthreads SO questions related to these answers. We classified the questions based on their formulation and technical content, in order to investigate RQ1. 
%Afterwards, we uniformly ranked all used and unused answers in each discussion thread in the descending order of scores, ages, code lengths, and text lengths. We analyzed the percentile ranks~\cite{Roscoe75} of answers to investigate RQ2.
%Afterwards, we manually inspected code snippets in referenced answers to categorize them based on the code content, and to study the distribution of answers among different categories for RQ2.
%Finally, we manually compared the content of reused answers and related Java files, to explore RQ3. 
%to decide how developers reused answers and to explore RQ3.

Our study revealed in total \totalthreads SO discussion threads referenced by \totalJava Java files, which belong to \totalproj GitHub projects.
Most of the threads (i.e., 78\%) are about solutions to coding problems: question askers describe their software requirements and answer providers offer programs to satisfy those requirements (RQ1). %14\% of the threads are about optimizations. Namely, askers describe their software requirements, provide some their own coding solutions, and seek for faster or shorter programs that outperform their solutions. Only 3\% of the threads are about debugging, where answerers help askers to debug code. 
The studied \totalthreads threads have in total \totalanswer answers. 
Using our hybrid approach, we found \totalref answers reused by GitHub codebases. 
%, with some of the answers reused by multiple codebases simultaneously. 
When ranking all answers in each thread, we observed that the reused answers have statistically higher scores, older ages, larger code, and more text (RQ2). Among the \totalentry scenarios of answer reuse, we found fully identical code snippets between GitHub and SO in only 40 scenarios. Most developers reused SO answers by applying changes to the suggested code, or creating code from scratch based on the ideas described by those answers.

%. Namely, the percentile rank of the median score among all referenced answers is X\%, while the median's percentile rank of unreferenced answers is Y\%. Compared with unreferenced answers, the referenced ones are usually older (A\% vs.~B\%), have larger code (C\% vs.~D\%), and contain more textual description (E\% vs.~F\%). 
%\todo{description on code snippets and code reuses}
In this paper, we made the following research contributions:

\begin{itemize}
\item We created a new approach to identify SO answers reused in open-source Java projects. By combining clone detection, keyword-based search, and manual inspection, this approach was able to identify reused answers effectively. 
\item We did a novel empirical study to characterize the SO answers reused by open-source projects on GitHub. Our study characterized the reused answers from unique angles like the discussion topics, the content, post ages, and post scores. No prior studies considered these aspects.
\item We defined a taxonomy of patterns to describe developers' answer-reuse practices. No prior work categorizes developers' reuse practices in such a comprehensive way.
\end{itemize} 
In the 
%remaining part of our paper, 
following sections, 
we will introduce the background (Section~\ref{sec:background}), our data collection process (Section~\ref{sec:method}), and experiments (Section~\ref{sec:result}).
We open-sourced our material (i.e., program and data) %our materials at
at 
%the website 
\underline{\url{https://figshare.com/articles/dataset/So-gitexperiment/20425839}}.
%, we open-sourced our program and data.

%(with top-3 ranks), middle ages (with 15$^{th}$--25$^{th}$ ranks), larger code (with top-3 ranks), and longer textual description (with top-4 ranks). 
%\vspace{-.5em}
\section{Background}\label{sec:background}
This section introduces the content of a typical StackOverflow discussion thread; it also shows how a GitHub project references a discussion thread and reuses one of the answers.

\begin{figure}
\includegraphics[width=\linewidth]{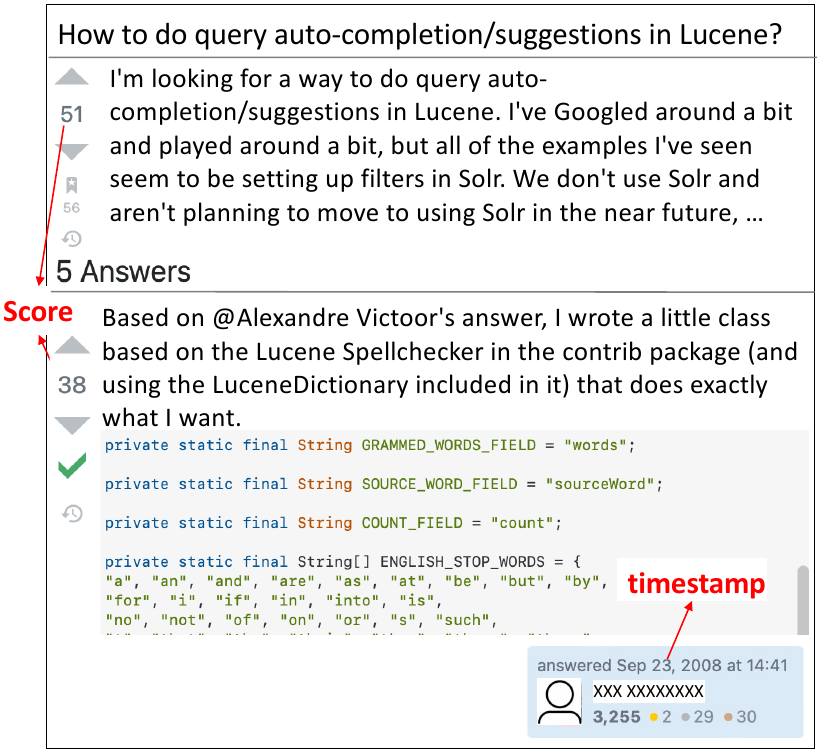}
\vspace{-2.5em}
\caption{An exemplar SO discussion thread that contains one question post and multiple answer posts~\cite{auto-completion}}\label{fig:so-discussion-thread}
\vspace{-1em}
\end{figure}
\begin{figure}
\includegraphics[width=\linewidth]{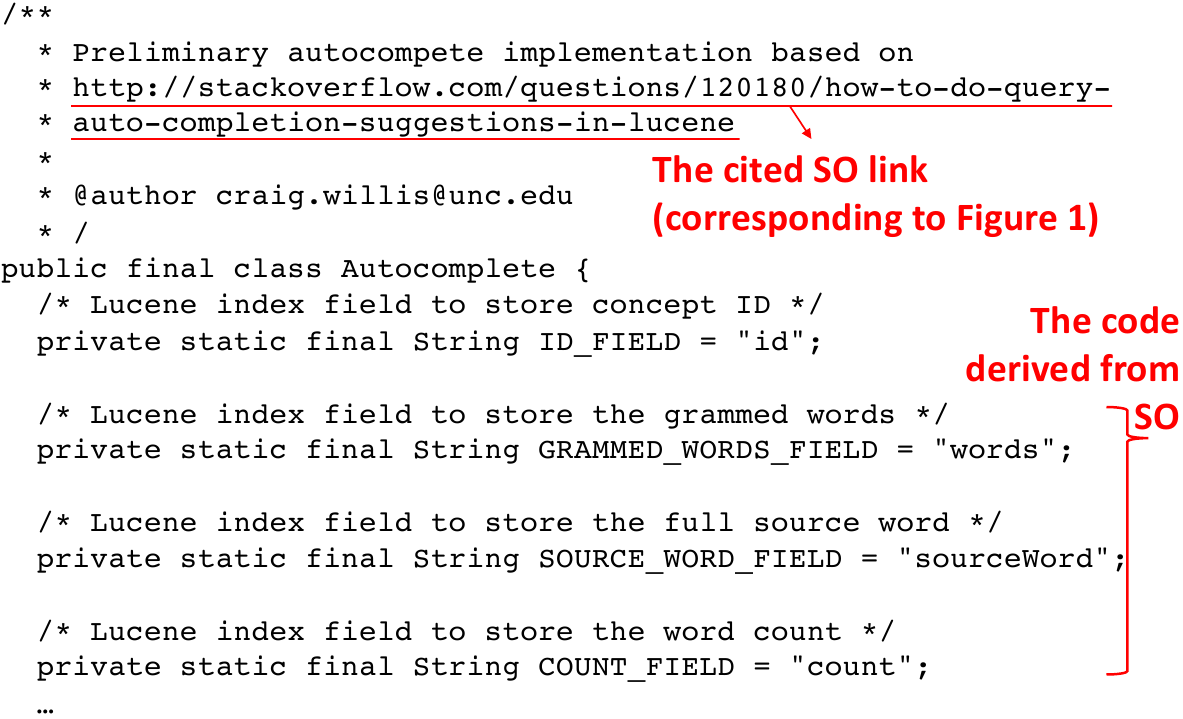}
\vspace{-2.5em}
\caption{An exemplar Java file on GitHub that cites an SO thread and reuses some of the answer code~\cite{github-autocomplete}}\label{fig:github-code-reuse}
\vspace{-1em}
\end{figure}

As shown in Figure~\ref{fig:so-discussion-thread}, 
an SO discussion thread has one \textbf{question} and zero, one, or multiple \textbf{answers} to that question. After a user posts a question or answer, other users can vote for or against that post. 
%Each user's \textbf{reputation} is determined by the votes their posts received. 
The \textbf{score} is decided by the up- and down- votes received by a post. Each post has a \textbf{timestamp} to show when it was created. 

StackOverflow assigns each question or answer a unique \textbf{ID number}, while the question ID is also treated as the \textbf{unique ID of its discussion thread}. A typical \emph{SO question link} has the format \codefont{https://stackoverflow.com/questions/} \codefont{question\_ID/question\_title}, 
while an \emph{answer URL} often has the format 
\codefont{https://stackoverflow.com/questions/} \codefont{question\_ID/question\_title/answer \_ID\#answer\_ID}. 

Note that the answer URL contains IDs of both the question and answer. 
The question and answer links mentioned above can be used to retrieve their corresponding discussion threads from StackOverflow. 
For instance, developers can retrieve the discussion thread shown in Figure~\ref{fig:so-discussion-thread} via the question URL 

``\url{https://stackoverflow.com/questions/120180/how-to-do-query-auto-completion-suggestions-in-lucene}'', 

\noindent
or the URL of any answer it contains, such as  

``\url{https://stackoverflow.com/questions/120180/how-to-do-query-auto-completion-suggestions-in-lucene/121456#121456}''. 

\noindent
Here ``120180'' is the question's ID% as well as discussion-thread ID
; ``121456'' is the answer's ID. In the datasets related to StackOverflow (e.g., Google BigQuery~\cite{bigquery} and StackOverflow data dump~\cite{datadump}), 
 a question/thread ID is often used as the \textbf{parent ID} of some answers, showing that those answers respond to the given question.

%for the exemplar discussion thread shown in Figure~\ref{fig:so-discussion-thread}, developers can retrieve it via the question URL 
%When multiple answers are available for an SO question, the asker may take the best one as \textbf{accepted answer} and mark it with  ``\textcolor{green}{\cmark}''. 

Figure~\ref{fig:github-code-reuse} shows a Java file on GitHub that references/cites the above-mentioned SO thread via its question link. Comparing the code content of Java file and the first answer in the thread, we can find common code that indicates developers' answer-reuse practice. 
%In software development, developers may cite the URL of an SO discussion thread (i.e., an SO question) in codebases, and reuse some of the answers mentioned in that thread (see Figure~\ref{fig:github-code-reuse}). 
Our research focuses on (1) the Java files from GitHub that explicitly reference URLs of SO questions or answers, and (2) the SO threads retrieved via those URLs. 

\section{Data Collection}
\label{sec:method}

To create a dataset of SO answers reused by GitHub projects, we used Google BigQuery~\cite{bigquery} to crawl for SO links that are mentioned by Java files on GitHub (Section~\ref{sec:crawl}), and applied a new hybrid approach to identify SO answers reused by developers (Section~\ref{sec:reuse}). 
%Based on the information of reused answers together with related discussion threads, we investigated 
%we performed a series of analysis to investigate  
%RQ1--RQ3 (Section~\ref{sec:analysis}). 

\vspace{-.5em}
\subsection{Data Crawling}\label{sec:crawl}
Google BigQuery is a serverless data warehouse that enables scalable data analysis. It is a Platform as a Service (PaaS) that supports SQL queries. 
 %querying using SQL.
It hosts datasets like GitHub project data and SO post information, and supports data queries across multiple datasets. Our research focuses on (1) the Java files on GitHub citing any SO links and (2) the SO posts related to cited links. Therefore, BigQuery satisfies our need.
%WHERE parent_id IS NOT NULL
%  AND (
%    SELECT COUNT(*)
%    FROM
%      'bigquery-public-data.stackoverflow.posts_answers' AS t2
%    WHERE a.parent_id = t2.parent_id

As shown in Listing~\ref{lst:sql}, we defined an SQL query for BigQuery. The query performs three major tasks. First, it retrieves Java files from GitHub that cite SO links and records those links (see lines 5--9). Second, treating each recorded link as the URL of a discussion thread, it retrieves all answer posts from SO belonging to that thread (see lines 1--10). Third, it orders retrieved answers based on the discussion-thread ID (i.e., \codefont{parent\_id}), and limits the total number of retrieved answers to 30,000 (see lines 11--12).
We set the upper bound to 30,000 for two reasons. First, our free user account with  BigQuery limits the number of SQL queries to execute each month, and the amount of data processed by each query. Second, our later data analysis involves manual inspection. Even if BigQuery can return all records satisfying the query, our manual inspection is  unscalable and can only analyze a subset. After trying different limit settings, we found 30,000 to be a reasonable number to handle.

\lstset{
numbers=left,
basicstyle=\scriptsize,
numberstyle=\scriptsize,
breaklines=true,
numbersep=1.5pt,
%escapeinside={(*}{*)},
frame=tb}
\begin{lstlisting}[language=sql, caption=Our SQL query to crawl GitHub and StackOverflow,label=lst:sql]
SELECT a.id, title, body, content, parent_id, favorite_count favs, view_count views,   score, accepted_answer_id, post_type_id, sample_repo_name, sample_path
FROM
  'bigquery-public-data.stackoverflow.posts_answers' a
INNER JOIN (
  SELECT
    CAST(REGEXP_EXTRACT(content, r'stackoverflow.com/questions/([0-9]+)/') AS INT64) id, sample_repo_name, sample_path, content
  FROM `fh-bigquery.github_extracts.contents_java`
  WHERE
    content LIKE '%stackoverflow.com/questions/%') b
ON a.parent_id=b.id 
ORDER BY parent_id
LIMIT 30000
\end{lstlisting}
\vspace{-.5em}

Among the records retrieved from Google BigQuery, we realized that no creation date of posts is included, and the \codefont{body} field does not always include the complete content of answer posts. As our later analysis heavily depends on the availability of posts' creation date and answer content, we conducted additional crawling on StackOverflow data dump~\cite{datadump} to acquire that information. 
In summary, our data crawling obtained 30,000 answer posts. These posts contain duplicates because when the same question ID is referenced multiple times, all of its answers are retrieved again and again. 
%which contains duplicates because when the same answer is referenced multiple times
%acquire the complete content of all answer posts.
%After our data crawling, we acquired 30,000 answer posts, 
After removing duplicates, we identified 322 unique SO discussion threads to cover all crawled answers. These threads are referenced by 1,254 unique Java files from 1,063 GitHub projects. %\textcolor{red}{While GitHub searches indicate a larger pool of around 108K Java files referencing OS threads as of 2024, statistical sampling methods were applied to select the dataset. The sample size strikes a balance between comprehensiveness and manageability. }

\vspace{-.5em}
\subsection{Detection of Answer Reuse}\label{sec:reuse}
As mentioned in Section~\ref{sec:background}, a Java file may reference an SO question and/or answer link. If an answer link is referenced, detecting answer reuse is simple: we only need to compare the Java file with cited answer to comprehend the reuse practice. However, if no answer is referenced and there is only one question link, detecting answer reuse is harder. As a question often corresponds to multiple answers, it is not always easy to decide which answer is more relevant. We believe that when developers reuse SO answers, they may reuse the code snippets mentioned in those answers; thus, we can find reused answers by detecting code clones (i.e., similar code) between each answer and the Java file. 
Based on this insight, 
We created a three-step hybrid approach to detect reused answers. 

\subsubsection{Step 1: Detecting reused answers via clone detection}
When a Java file $f$ references an SO question link, we located all answers to that question. To facilitate discussion, we use $A$ to denote the located answer set. From each answer $a_i\in A$, we extracted all code snippets and stored them in a single Java file $j_i$. We denote all Java files synthesized in this way with $J$. Afterwards, we applied PMD Copy/Paste Detector (CPD)~\cite{pmd}---a clone detection tool---to $f$ and $J$, in order to find code clones between the Java file and answer code.
We chose PMD because it is publicly available, easy to use, and often applicable given two files for comparison. 
When comparing Java programs, PMD treats source code as plain text and adopts a string-matching algorithm~\cite{karp-rabin} to compare the hash values of strings. It does not matter if some synthesized Java files are incompilable or have lexical/syntactic errors, because PMD does not observe program syntax or semantics anyway. 

PMD has a parameter \codefont{minimumTokenCount} to specify the minimum number of tokens contained by a reported clone. This parameter controls trade-offs between the effectiveness of clone detection and the usefulness of detected clones. Namely, if the parameter value is too small (e.g., 2), PMD can find a lot of clones; however, many of the clones may share as few as two tokens and seem to be accidental code overlap instead of meaningful code reuse. Meanwhile, if the value is too large (e.g., 500), PMD may only report few clones while each reported clone pair  
 %of the reported clone pairs 
 shows strong evidence of code reuse. %By default, PMD set this parameter to 100. However, we noticed that SO answers often contain small code snippets with lengths less than 100. We should set the parameter to a much smaller number.%We should not blindly use the default parameter setting. 

To properly configure \codefont{minimumTokenCount}, we conducted a preliminary study by experimenting PMD with different parameter values. Specifically, we randomly picked 30 pairs of $\langle f, a_i\rangle$. Here, $f$ represents a Java file from GitHub referencing a discussion thread, and $a_i$ represents a Java file synthesized from an answer of that thread. 
We tuned \codefont{minimumTokenCount} from 5 to 25, with 5 increments. For each parameter setting, we applied PMD to the 30 $\langle f, a_i \rangle$ file pairs for clone detection. Then we manually inspected results to see which setting achieves a better trade-off between the number and quality of reported clones. For simplicity, if multiple clones are reported for a given file pair, we checked only the first clone pair to assess whether the code has meaningful similarity or just random content overlap. With ``\textbf{random/accidental content overlap}'', we mean a few common tokens shared between two totally different statements.
In this experiment, for each setting, we examined at most 30 clone pairs when PMD reported clones for all file pairs.

\begin{table}
\caption{Number of clones PMD reported when {\tt minimumTokenCount} was set differently}
\label{tab:param1}
\vspace{-1.3em}
\scriptsize
\centering
\begin{tabular}{p{3.cm}| r| r| r| r| r}
\toprule
{\tt minimumTokenCount =}&\textbf{5} &\textbf{10 } &\textbf{15} &\textbf{20 } &\textbf{25 }
\\  \toprule
\textbf{\# of clone pairs examined} & 30 &22 &17 &12 &10\\ \hline
\textbf{\# of true positives} &12 &12 &12&12&10\\ \hline
\textbf{Precision} &40\%&55\%&71\% &100\%&100\%\\ \bottomrule
\end{tabular}
\vspace{-2.5em}
\end{table}

As shown in Table~\ref{tab:param1}, when \codefont{minimumTokenCount} increases, the total number of clone pairs examined decreases. This is because when more common tokens are required between clones, PMD reported clones for fewer file pairs.
Among the manually checked clone pairs, 
%we also counted how many pairs present meaningful code duplicates instead of accidental code overlaps. W
we identified 12 true clone pairs when \codefont{minimumTokenCount} was 5, 10, 15, or 20. However, only 10 true clone pairs were identified when the parameter was 25. %This is understandable because when the parameter becomes larger, fewer clone pairs can satisfy the requirement and some true clones may get missed as a result. 
Finally, we computed the precision rate for PMD by dividing the number of true positives/clones with the total number of clone pairs examined. %As expected,
When the parameter increases, the precision rate increases or remains. Considering both the total number of true clones revealed and precision rate, we found \codefont{minimumTokenCount=20} to be a better setting than the others.

We did another experiment to further tune the parameter from 17 to 20 tokens, with 1 increments. We wanted to explore a setting that outperforms 20 by achieving a better trade-off between the number of true positives and the precision rate. 
%Due to the space limit, we do not show our results here. In summary, 
We found 18, 19, and 20 to produce equally good experiment results. Therefore, by default, we set \codefont{minimumTokenCount=18}, to retrieve as many true positives as possible while ensuring high precision. With this setting, PMD reported clones for 1,526 $\langle f, a_i\rangle$ pairs in our dataset, which pairs are considered as candidates to show answer reuse.
%We consider those pairs as candidates to show answer reuse.  

\subsubsection{Step 2: Detecting reused answers via keyword-based search}
When a Java file $f$ references an SO answer link, we consider the answer to be potentially reused by $f$. Based on our experience, although Step 1 is effective in finding candidates, it can miss reused answers in three kinds of scenarios. First, there is no code mentioned in an answer, but developers reused that answer by digesting the idea and writing code accordingly. Second, developers got inspired by code mentioned in an answer, but wrote totally different code. 
%but wrote Java files that have totally different content. 
%that are totally different from the answer code based on inspirations. 
Third, due to some unknown implementation issues, PMD fails to identify code clones between certain similar code.

To find the reused answers potentially missed by PMD, we decided to rely on SO answer links explicitly referenced by Java files. Specifically, we extracted the IDs of 30,000 retrieved answers, and used them as keywords to search among the 1,254 Java files we crawled (see Section~\ref{sec:crawl}). If a Java file contains any of the answer IDs, the file is considered to potentially reuse that answer. With this approach, we found 89 Java files citing the answers under analysis.

\subsubsection{Step 3: Manual inspection to refine detected answer reuses} Among the $\langle f, a_i\rangle$ pairs that PMD reported to have clones, not every pair implies an actual answer reuse. 
For instance, if the latest editing date of a Java file $f$ is in 2015 but $f$ has code similarity with an SO answer $a_i$ posted in 2021, $f$ can never reuse $a_i$ as the answer did not exist in 2015. To identify actual answer reuses, we used the following four criteria to remove unpromising pairs:

\begin{itemize}
\item \textbf{Date Comparison:} If $\langle f, a_i\rangle$ has the creation date of $a_i$ later than the latest editing date of $f$, the pair does not imply answer reuse as developers could not refer to a nonexistent answer when editing their file. %Thus, we removed such pairs.
\item \textbf{Similarity Comparison:} If (1) $f$ is reported to have clones with several answers from the same thread and (2) those answers share code, 
we consider the answer that has the highest code similarity with $f$ as the reused answer. We believe that when reusing answer code, developers are more likely to focus on one answer instead of reusing multiple answers simultaneously. %merging the code from different answers.
\item \textbf{ID Comparison:} If $f$ explicitly cites an answer link and gets reported to have clones with several answers from the same thread, we consider the cited answer to be reused and treat all other answers as false positives. This is because developers are likely to reference the answers that help them most, giving credits to the reused answers.
\item \textbf{Availability Checking:} If $f$ is not available because developers recently removed that file or the whole project, we cannot compare the file with any candidate answers it may reuse. Thus, 
%In such scenarios, 
we removed all pairs containing $f$.
%In such scenarios, even though some unreferenced answers have code similarity with developers' code, the similarity is just accidental. 
\end{itemize}
%With the criteria mentioned above, we manually examined all reported file pairs to decide which Java files actually reuse what answers. This process was very time-consuming and error-prone. 
Our manual analysis of clone pairs was very time-consuming and error-prone. 
To reduce human errors, two authors separately checked reported file pairs, and then held meetings to go over the list. They compared their manual inspection results; whenever the results diverged, they discussed comprehensively to reach a consensus.

After identifying answer reuses based on PMD results, we further examined the 89 Java files found via keyword-based search in Step 2. We used two major criteria to decide whether a file's answer reference should be considered as answer reuse:

\begin{itemize}
\item \textbf{Duplicated Entries:} Suppose that $f$ explicitly cites $a_i$, while the pair $\langle f, a_i \rangle$ is already included into our dataset of answer reuse due to clone detection. In such scenarios, the Java file's answer reference is not added redundantly. 
%we do not add any entry for the Java files' answer references. 
\item \textbf{Availability Checking:} If $f$ is unavailable because developers recently removed it or the whole project, we cannot compare the file with any answer it references. 
Thus, the Java file's answer reference is not added to our dataset.
%In such scenarios, the Java file's answer reference is not added to our dataset of answer reuse.
%\item \textbf{Self Reference:} Suppose that $f$ cites $a_i$, while the authors of $f$ and $a_i$ have the same user ID. It is likely that the Java developer cites his/her own SO answer, %In such cases, so we do not consider the file's reference as answer reuse.
%such Java files' answer references as answer reuse.
\end{itemize} 

\begin{figure*}
\centering
\includegraphics[width=.85\linewidth]{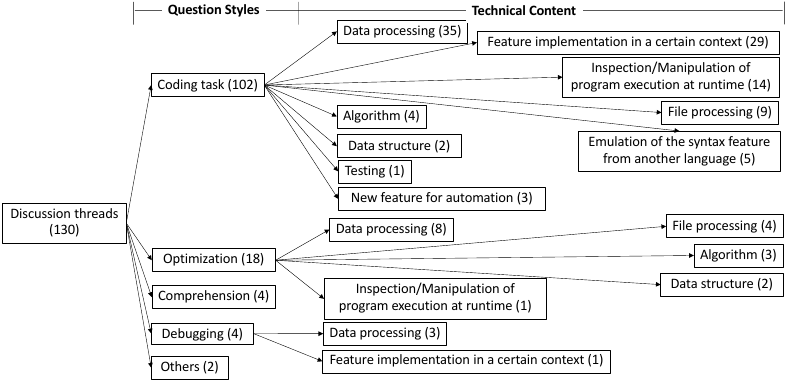}
\vspace{-1.3em}
%\caption{The topics we recognized in the \totalthreads discussion threads}
\caption{Our taxonomy of SO threads based on the discussion topics}
\label{fig:topics}
\vspace{-1em}
\end{figure*}
%With the criteria mentioned above, we manually inspected all reported Java-SO references to decide which Java files reuse what answers. To avoid human bias, the first and last authors separately examined the reported references. 
Two authors followed the criteria mentioned above to separately examine the reported references. They then compared and discussed results until reaching a consensus.

By applying the hybrid approach mentioned above, we got a refined dataset to include 
 \totalthreads discussion threads, which contain \totalanswer answers in total. Among those answers, \totalref answers were found to be reused, and \totalnonref answers were unused. The \totalref answers were reused \totalentry times; \totalJava Java files from \totalproj unique projects reused those answers. 
 Our later experiments (see Section~\ref{sec:result}) will focus on this refined dataset.
 % together with the Java files involved in answer reuse.

%\section{Experiment Results}\label{sec:result}
\section{Experiments}\label{sec:result}

This section presents our investigation for the three research questions RQ1--RQ3. For each RQ, we will first introduce the study method and then describe our experiment results.
%our experiment results. 

\vspace{-.5em}
\subsection{RQ1: Topics of Discussion Threads}

\paragraph*{\textbf{Study Method}}
RQ1 captures the focus of developers when they reused answers. %RQ1 intends to characterize the topics for discussion threads that have answers reused by developers, in order to capture developers' concerns when they search SO for technical assistance. 
We performed a lightweight open coding-like process~\cite{Seaman99} to categorize the topics discussed in \totalthreads threads. Specifically, while the last author (a professor with SE expertise) manually inspected all \totalthreads questions in threads, she extracted or summarized keywords to characterize each question in terms of the question style and technical content. Then she identified the commonality between questions based on keywords, defined categories accordingly, and revisited all questions to decide whether the categories were comprehensive enough to cover all questions. The author defined and refined categories iteratively, until each question was mapped to a category and the category list was stable. 
Next, the first author manually checked the categorization results, and discussed with the last author if any classification seemed problematic.

\paragraph*{\textbf{Results}}
Figure~\ref{fig:topics} presents our classification results of SO threads. In terms of the question style, we identified 5 categories in the \totalthreads discussion threads: coding task, optimization, comprehension, debugging, and others. \emph{Coding task} means that askers describe software requirements (e.g., how to validate an XML file against an XSD file~\cite{so-15732}) and seek for code solutions, while answerers offer code to implement those requirements. 
%\textbf{Optimization} is similar to the coding-task category in a sense that askers also describe their software requirements, but is different because askers offers their initial implementation satisfying certain requirements  
\emph{Optimization} means that askers provide initial programs satisfying certain requirements, looking for better programs that have 
%means that question askers describe their programs for specific software requirements, and look for better programs that have
 either easier implementation, lower runtime overheads, or less platform-specific dependency~\cite{so-101439}. 
An exemplar thread of this category is about the most efficient way to implement an integer-based power function pow(int, int)~\cite{so-101439}.
Optimization is different from the coding-task category, as askers provide initial code implementation. 

\emph{Comprehension} is about clarification or comparison of concepts or terms. For instance, an SO thread compares two Java APIs: StringBuilder versus StringBuffer~\cite{so-355089}. 
\emph{Debugging} means that askers present their erroneous programs and seek for debugging advices. An example of this category is about a strange OutOfMemory issue while loading an image to a Bitmap object~\cite{so-477572}. \emph{Others} captures the miscellaneous questions not covered by any category mentioned above. Specifically, one of the two questions is about SQLite command usage~\cite{so-525512}, and the other focuses on issues that developers should consider when overriding \codefont{equals(...)} and \codefont{hashCode()}~\cite{so-27581}.

Among the five major categories, we noticed that coding task dominates the discussion threads. It covers 102 of the \totalthreads threads. The second biggest category is optimization, which covers 18 threads. These observations imply that when developers reuse SO answers, they often focus on the answers that offer complete (or even optimized) programs to satisfy certain requirements. Surprisingly, comprehension and debugging separately cover only four threads. It means that although StackOverflow provides a platform for developers to discuss technical concepts or software bugs, developers rarely reused the answers related to concept comprehension or bug fixes. This may be because comprehension-related discussion is too general or abstract for developers to adopt in code, while debugging-related discussion is too specific or concrete for developers to reuse in their diverse circumstances. 

\vspace{0.2em}
\noindent\begin{tabular}{|p{8.4cm}|}
	\hline
	\textbf{Finding 1:} \emph{92\% (120/130) of discussion threads are about coding tasks or optimizations. The reused answers often provide complete (or even optimized) coding solutions for given software requirements.}
	\\
	\hline
\end{tabular}
%\vspace{0.2em}

In terms of the technical content, we identified 9 topics in the \totalthreads threads (see Figure~\ref{fig:topics}). \emph{Data processing} is about how to generate, handle, or transform data. For instance, a thread discusses how to do 3DES data encryption/decryption~\cite{so-20227}. Similarly, \emph{File processing} is about how to handle or compare  files. An exemplar thread is about optimized ways of counting lines in a file~\cite{so-453018}. Concerning 
\emph{Feature implementation in a certain context}, askers describe their software requirements in specialized circumstances (e.g., when using certain Java libraries), and seek for suggestions applicable to those circumstances. For instance, a thread discusses how to make a list with checkboxes in Java Swing~\cite{so-19766}. With \emph{Inspection/Manipulation of program execution at runtime}, we refer to discussions on how to programmatically inspect or manipulate execution status/environments at runtime, such as programmatically checking CPU and memory usage~\cite{so-74674}. 

\begin{figure*}
\centering
\includegraphics[width=0.84\linewidth]{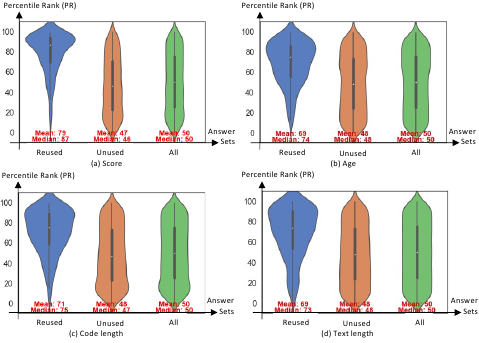}
\vspace{-1.5em}
\caption{The PR-comparison among reused answers, unused answers, and all answers of the \totalthreads threads}\label{fig:pr-comparison}
\vspace{-1.5em}
\end{figure*}

\emph{Algorithm} focuses on the design and implementation of algorithms or mathematical computation, such as calculating the distance between two latitude-longitude points~\cite{so-27928}. \emph{Emulation of the syntax feature from another language} focuses on how to emulate certain syntax features offered by languages other than Java, such as the as-operator of C\#~\cite{so-148828}. \emph{Data structure} covers discussions on defining customized data structures (e.g., an LRU cache~\cite{so-221525}).
\emph{Testing} is about defining test cases. 
For instance, one thread discusses how to test methods that call System.exit()~\cite{so-309396}. 
\emph{New feature for automation} describes the threads that discuss rare feature implementation, which is not covered by any of the topics mentioned above. For instance, a thread discusses how to play sound in Java~\cite{so-26305}. 

Among the five categories mentioned above, the dominant category coding task covers all nine topics. In particular, within the 102 discussion threads, 35 threads focus on data processing, 29 threads discuss feature implementation in a certain context, and 14 threads are about the inspection or manipulation of program execution at runtime. The optimization category covers five topics: data processing, file processing, algorithm, data structure, and inspection/manipulation of program execution at runtime. The debugging category covers even fewer topics: data processing and feature implementation in a certain context, mainly because there are a lot fewer threads covered by this category.

Among the nine topics, the most popular four topics are: data processing, feature implementation in a certain context, inspection/manipulation of program execution at runtime, and file processing. Each of the topics separately covers 46, 30, 15, and 13 threads. %Our observations imply that when answerers respond to questions of these four topics, their answers are more likely to be reused by developers. 

\vspace{0.2em}
\noindent\begin{tabular}{|p{8.4cm}|}
	\hline
	\textbf{Finding 2:} \emph{80\% (104/130) of threads are on the topics of data processing, feature implementation in a certain context, inspection/manipulation of program execution at runtime, and file processing.% When answerers respond to such questions, their answers are more likely to be reused.
 }
	\\
	\hline
\end{tabular}
%\vspace{0.2em}

%\vspace{-.5em}
\vspace{-.5em}
\subsection{RQ2: Characteristics of Reused SO Answers}
\paragraph*{\textbf{Study Method}} 
RQ2 explores which of the following characteristics can effectively differentiate reused answers from unused ones: post scores, post ages, lengths of answer code, and lengths of the text (i.e., code + explanation). 
 A challenge in comparing the two sets (reused vs.~unused answers) is that we cannot na\"ively compare the measured absolute values, 
%values for any metric between the two answer sets. 
because the measured values from different threads can vary a lot. For instance, one discussion thread has answer scores vary within [1, 5], while another thread has the scores in [10, 5000]. If we na\"ively extract scores (or other values) and compare the distributions, it is almost impossible for us to see any meaningful difference between sets.

To overcome the challenge, we defined an approach to compare the two sets based on relative ranks of answers in each thread.
% instead of the absolute values measured for any metric. 
Specifically, we wrote scripts to automatically sort the answers of each discussion thread based on four alternative metrics: post scores, post ages, code lengths, and text lengths. The post scores were directly extracted from the data crawled by BigQuery (see Section~\ref{sec:crawl}). The post ages were computed based on the date difference between each post's creation and our experiment. 
%creation dates of posts extracted from StackOverflow (SO) data dump. 
%crawled by BigQuery.
 Both code lengths and text lengths were based on the complete answer content extracted from SO data dump. In particular, code length is the character count of each answer code, while the text length is the character count of each answer. In our experiment, the maximum and minimum text lengths are separately 24,367 and 29. 

For each sorting task, we ranked answer posts in descending order of measured values, because we intended to explore whether reused answers were usually ranked differently from unused answers.
Based on the ranking results, we computed a percentile rank (PR)~\cite{Roscoe75} for each answer as below: 

\vspace{-1em}
\begin{equation}
PR(v) = \frac{CF(v) - 0.5 \times F(v)}{N} \times 100
\end{equation} 
PR is within [0, 100]. 
CF means \emph{cumulative frequency}---the count of all values less than or equal to the value of interest $v$. F is the \emph{frequency} of $v$. N is the total number of answers in the ranked list. For instance, if we have a ranked list \{5, 5, 4, 3, 2, 1, 0\}. Then $PR(5) = \frac{7-0.5\times2}{7}\times 100=88$. As another example, when posts are ranked in descending order of ages, the oldest and newest posts separately get the highest and lowest PR values.  

By mapping concrete measurements to PR values, we compared the relative rank distributions of reused and unused answers within their separate discussion threads. 
In data analytics, violin plots visualize the distributions of numerical data~\cite{violin}. They show not only summary statistics (e.g., median and interquartile range), but also the density of each variable. We used violin plots to visualize the PR distributions of reused answers, unused answers, and all answers of the \totalthreads threads.  
We also conducted Mann-Whitney U test~\cite{mann-whitney} 
%Kruskal-Wallis H test~\cite{kruskal-wallis}
 to check whether the distributions of reused and unused answers present statistically significant differences.

\paragraph*{\textbf{Results}} 
Figure~\ref{fig:pr-comparison} visualizes the PR distributions of three answer sets based on four measurements: score, age, code length, and text length. 
The \emph{Reused} set clusters the \totalref answers reused by \totalJava Java files.  %we identified in Section~\ref{sec:method}. 
The \emph{Unused} set clusters the remaining \totalnonref answers included by the  \totalthreads threads. 
%, but not found to be reused by any Java file we examined. 
The \emph{All} set includes all \totalanswer answers. %To draw each plot in Figure~\ref{fig:pr-comparison}, we computed the PR values of answers separately based on answers' scores, ages, code lengths, and text lengths (Section~\ref{sec:rq2}).

As shown in Figure~\ref{fig:pr-comparison}, reused answers usually have higher scores, older ages, larger code, and longer text than unused ones. In terms of scores (Figure~\ref{fig:pr-comparison} (a)), the PR values of reused answers have the mean  as 79 and median as 87. The upper quartile (i.e., the value under which 75\% of data points are found) is 94, and the lower quartile (i.e., the value under which 25\% of data points are found) is 69. The data peak is near the upper quartile---94.  
On the other hand, the PR values of unused answers have the mean as 47 and  median as 46. %The upper quartile is 71 while the lower quartile is 22.
The plots for both unused and all answers have no obvious peak. One possible reason to explain our observations is that scores reflect the quality of answers. 
As developers often strive to reuse the answers with highest quality, it is unsurprising to see that reused answers often have much better scores than unused ones. 
Our observations imply that when developers hesitate to choose answers from a thread, they can pay more attention to the few with highest scores. 

In terms of ages (Figure~\ref{fig:pr-comparison} (b)), the PR values of reused answers have the mean as 69 and median as 74.
%; the upper and lower quartiles are separately 85 and 55. 
%Most of the data points fall into [40, 100]; the chart has a long tail falling into the range [0, 40]. Meanwhile, 
The PR values of unused answers have both the mean and median as 48. 
%; the upper and lower quartiles are 73 and 24 respectively. 
%The data points of unused answers spread more evenly within the whole range [0, 100], with no obvious data peak.
One possible reason to explain the phenomena is our data-crawling process. Because we crawled GitHub projects to reveal answer-reuse practices, many of the practices reflected by source code were actually conducted months or years ago. Thus, the answers reused are typically older. 
In terms of code lengths and text lengths (Figure~\ref{fig:pr-comparison} (c) \& Figure~\ref{fig:pr-comparison} (d)), we observed similar phenomena. 
%We observed similar phenomena in the graphs for code lengths and text lengths (Figure~\ref{fig:pr-comparison} (c) \& Figure~\ref{fig:pr-comparison} (d)).
%the newest; instead, they are often older. 
%These observations indicate that when developers hesitate to choose answers for reuse from a thread, they can focus more on the candidate answers that are older than 40\% of the answers.
%In the graphs for code lengths and text lengths (Figure~\ref{fig:pr-comparison} (c) \& Figure~\ref{fig:pr-comparison} (d)), we observed similar phenomena to the ones described for the age graph Figure~\ref{fig:pr-comparison} (b). 
In both graphs, reused answers have 73--75 medians, while unused answers have 47--48 medians. 
%, 89--90 upper percentiles, and 54--58 lower percentiles. %Most data points fall into the range [40, 100]. Meanwhile, 
%The PR values of unused answers have 47--48 medians, 73 upper percentile, and 23--24 lower percentiles.
One possible reason to explain our observations is that the longer code or text an answer has, the easier it is for developers to digest the idea and reuse that answer. 
Our observations imply that when developers hesitate to choose answers for reuse, they can focus more on the answers with larger code or longer description.  
%either larger code or longer textual description. 
%Additionally, if answerers want to create more reusable answers, 
%want to attract more attention from developers,  
%who may potentially reuse answers, 
%they can provide answers that have longer code or text. 
 
The Mann-Whitney U test we performed shows %Kruskal-Wallis H test 
%to check whether reused and unused answers present different PR distributions. We observed 
$p<<0.05$ for all measurements, meaning that the reused and unused answers have statistically significant differences in terms of their PR ranks by score, age, code length, and text length.

\begin{table*}
\caption{The identified types of developers' answer-reuse practices among the \totalentry reuse scenarios}\label{tab:reuse-types}\vspace{-1.3em}
\scriptsize
\begin{tabular}{l|p{0.5cm}| p{4cm}| p{9.cm} |r |r}
\toprule
\multicolumn{2}{c|}{\textbf{Idx}} &\textbf{Name} &\textbf{Definition} &\textbf{Count} &\textbf{Percentage} \\ \toprule
\multicolumn{2}{c|}{C1} & Exact copy & Developers copy and paste \emph{all} code, without any modification. & 40 &9\%\\ \hline
\multicolumn{2}{c|}{C2} & Copy with cosmetic modification & Developers copy and paste \emph{all} code, and apply minor modification without changing the program structure. The minor modifications are limited to updates to identifiers and literals.& 94 &22\%\\ \hline
\multirow{20}{*}{C3} 
&C3.1 &Reuse all with statement-level updates only&Developers copy and paste \emph{all} code, and update content of some statements.&10&2\% \\ \cline{2-6}
&C3.2 &Reuse all with structure changes only& Developers copy and paste \emph{all} code; they modify the program structure by inserting, deleting, reordering, or combining statements. &13&3\%\\ \cline{2-6}
&C3.3&Reuse all with both statement-level updates and structure changes & Developers copy and paste \emph{all} code; they update content of some statements; they also revise the program structure by inserting, deleting, reordering, or combining statements.&57&13\%\\ \cline{2-6}
&C3.4&Reuse most without change & Developers copy and paste \emph{most} code, without applying any change. &13&3\%\\ \cline{2-6}
&C3.5&Reuse most with statement-level updates only& Developers copy and paste \emph{most} code, and update content of some statements.&21&5\% \\ \cline{2-6}
&C3.6&Reuse most with structure changes only & Developers copy and paste \emph{most} code; they modify the program structure by inserting, deleting, reordering, or combining statements. &12&3\%\\ \cline{2-6} 
&C3.7&Reuse most with both statement-level updates and structural changes &Developers copy and paste \emph{most} code; they update content of some statements; they also modify the program structure by inserting, deleting, reordering, or combining statements. &84&20\%\\ \cline{2-6}
&C3.8&Reuse some without change & Developers copy and paste \emph{some} code, without applying any change.&3&1\%\\ \cline{2-6}
&C3.9&Reuse some with statement-level updates only& Developers copy and paste \emph{some} code, and update content of certain statements.&19&4\% \\ \cline{2-6}
&C3.10&Reuse some with structure changes only & Developers copy and paste \emph{some} code; they modify the program structure by inserting, deleting, reordering, or combining statements.&2&*0\% \\ \cline{2-6} 
&C3.11&Reuse some with both statement-level updates and structural changes &Developers copy and paste \emph{some} code; they update content of certain statements; they also modify the program structure by inserting, deleting, reordering, or combining statements.&38&9\% \\  \hline
\multicolumn{2}{c|}{C4} &Converting ideas&Developers do not copy or paste any code. Instead, they write code from scratch based on the ideas shown by the code example(s), algorithm description, or image(s) in an answer. &7&2\% \\ \hline
\multicolumn{2}{c|}{C5} &Learning knowledge& Developers do not copy or paste any code. Instead, they write code based on the concepts or term definitions described in an SO answer.&17 &4\%\\
\bottomrule
\multicolumn{6}{l}{* 0\% is actually 0.47\%. The table shows 0\% because we rounded the measured value to the nearest integer.}\\ \bottomrule
\end{tabular}
\vspace{-2.5em}
\end{table*}

% create answers with larger code and longer textual description.
\vspace{0.5em}
\noindent\begin{tabular}{|p{8.4cm}|}
	\hline
	\textbf{Finding 3:} \emph{Compared with unused answers, reused ones have statistically higher scores, longer code, and longer text. 
	%This may be because developers often reuse answers with high quality. 
	Higher scores imply better quality; longer code and longer text help improve answers.}
	\\
	\hline
\end{tabular}

\vspace{-1.em}
\subsection{RQ3: Categorization of Reuse Practices}
\paragraph*{\textbf{Study Method}} 
RQ3 investigates how developers reused SO answers in their software projects. We took a lightweight open coding-like process to categorize developers' reuse practices demonstrated in \totalproj GitHub projects. Specifically, 2 authors separately checked \totalentry cases (i.e., code locations) where SO answers were reused, and defined keywords to summarize the similarity between Java files and SO answers. Then they held a meeting to compare summaries for different kinds of cases, discussed the rationale behind those summaries, and brainstormed a set of categories that are exclusive of each other but sufficient to cover all observed cases.
Next, the authors divided the \totalentry cases into 2 sets, and separately worked on a set to label the reuse practices based on the category set they agreed upon. Afterwards, they crosschecked the labels assigned by each other, and had discussions as needed to reach a consensus.

\vspace{-.5em}
\paragraph*{\textbf{Results}} 
Depending on the observed similarities between Java files and the SO answers they reused, we identified 5 major types (i.e., C1--C5) of answer-reuse practices in the crawled \totalentry reuse scenarios. As shown in Table~\ref{tab:reuse-types}, C1--C3 involve copying and pasting code, while C4--C5 do not copy or paste any code. 
In particular, C1 means exact copy: developers fully copy and paste code without modifying anything~\cite{so-1966151,github-1966151}. C2 means full copy with cosmetic modifications like identifier renaming and literal replacement~\cite{so-26318,github-26318}. C3 means copy with non-cosmetic modifications~\cite{so-27609,github-27609}. C4 means idea reuse: an answer illustrates  coding ideas via code examples, algorithm explanation, or images~\cite{so-196065,github-196065}; developers reuse the ideas without reusing any code. C5 means knowledge learning: an answer explains certain concepts or terms; developers digest the concepts/terms to independently work on their coding tasks~\cite{so-163539,github-163539}. 

Because C4 and C5 have no code reuse, we recognized the reuse practices for both categories based on the answer IDs cited by Java files and manual inspection (Section~\ref{sec:reuse}). Namely, 
 %answer reuse purely based on the cited answer IDs in GitHub code.
  If a Java file cites an answer and implements the algorithm described by that answer, we considered the answer reuse as C4; otherwise, it is of C5. Because C1--C3 have code reuse, we identified the reuse practices for these categories mainly based on clone detection and manual inspection. 

%Both C4 and C5 show answer reuse without code reuse: although there is no code similarity between SO answers and GitHub code, we identified answer reuses based on the answer IDs explicitly cited by GitHub code.
% If a cited answer describes the algorithm implemented in the citing GitHub code, the answer reuse belongs to C4; otherwise, it belongs to C5.

As shown in Table~\ref{tab:reuse-types}, developers created exact copies for answer code in only 9\% of cases; they modified and reused answer code in 85\% of cases, and did not reuse any answer code in the remaining 6\%. These numbers provide two insights. First, developers seldom used answer code as is, but customized code before reuse.
Second, code reuse is the most typical way that developers took when reusing answers. Thus, if answerers want to impact the coding practices of more developers, they'd better provide code examples in answers. 
Among C1--C3, C3 is the largest category, covering 63\% of scenarios. To better characterize developers' reuse practices in these scenarios, we defined 11 subcategories for C3 based on 3 criteria: 
%our code inspection results in three aspects:

\begin{enumerate}
\item \textbf{Did GitHub developers reuse all, most, or some of the answer code?} Given an answer, if a Java file has counterparts (i.e., similar or identical statements) for all statements in the answer code, we use ``all'' to describe the reuse level. Alternatively, if a Java file has counterparts for more than 50\% of the statements in answer code, we use ``most''; otherwise, we use ``some''.
\item \textbf{Did GitHub developers modify the content of any statement in the answer code?} If developers updated any statement in the answer code before reusing it, and if the updated code is similar to the original one, we use ``yes'' to mark the update. Otherwise, we use ``no''. 
\item \textbf{Did GitHub developers modify the program structure of answer code?} If developers added, deleted, reordered, or combined statements in the answer code, we use ``yes'' to mark the structure changes. Otherwise, we use ``no''. Structure change is orthogonal to statement update, because statement updates do not involve adding, deleting, reordering, or combining statements; instead, they are purely about minor changes in individual statements.
\end{enumerate} 

As shown in Table~\ref{tab:reuse-types}, C3.3, C3.7, and C3.11 are the three most popular ones among the 11 subcategories.
% they separately account for 13\%, 20\%, and 9\% of the overall \totalentry scenarios. 
Cases in these subcategories applied both statement-level updates and structure changes to the reused code. 
 There are six subcategories involving structure changes: C3.2, C3.3, C3.6, C3.7, C3.10, and C3.11; they account for 48\% in total. Meanwhile, there are six subcategories involving statement-level updates: C3.1, C3.3, C3.5, C3.7, C3.9, C3.11; they account for 53\% in total. Our observations imply that when developers applied non-cosmetic changes to copied code, they were more likely to apply statement-level updates than structure changes.

\vspace{0.2em}
\noindent\begin{tabular}{|p{8.4cm}|}
	\hline
	\textbf{Finding 4:} \emph{Most developers reused at least some of the answer code. 
	%Instead of copying and pasting code as is, 
	Developers often revised the code-to-reuse by applying statement-level updates and/or structure changes.}
	\\
	\hline
\end{tabular}

\vspace{-1em}
\section{Related Work}\label{sec:related}

The related work includes studies on SO, and studies on the relationship between SO and GitHub.

%of The Crowdsourced Knowledge Available
\vspace{-1.em}
\subsection{Studies on StackOverflow}
Researchers performed studies to characterize the crowd-sourced knowledge  on StackOverflow~\cite{Nasehi2012,Movshovitz-Attias2013,Gantayat2015,Honsel2015,meng2018secure,Zhang2018,Meng2019:icse}. 
Specifically,  
Movshovitz-Attias et al.~\cite{Movshovitz-Attias2013} analyzed the SO reputation system to identify the participation patterns of high and low reputation users.  
%They found that very high reputation users are the primary source of answers;  
%and especially of high quality answers; 
%the majority of questions were asked by low reputation users. 
Honsel et al.~\cite{Honsel2015} first interviewed five developers to identify the nine myths they believed, and then analyzed SO data to check those myths. 
%Their  analysis confirmed four of the myths (e.g., positively voted questions are more likely to get an answer), and busted the others (e.g., answers with too much text get worse votes). 
Gantayat et al.~\cite{Gantayat2015} studied the synergy between voting and acceptance of answers on SO, finding the accepted answers to be top-voted in 81\% of threads. 
Some researchers examined SO threads for specialized domains. 
%, to characterize content of the questions and answers. 
%either topics of the questions or quality of the answers. 
For instance, 
%Pinto et al.~\cite{Pinto2014} analyzed energy consumption-related posts to explore developers' concerns, the important aspects of energy consumption, and developers' solutions to improve energy efficiency. 
Zhang et al.~\cite{Zhang2018} studied the code examples of API usage, to reveal answers with API misuses.
% They observed that 31\% of analyzed SO posts may have 
%potential API misuses. 
%mined for energy consumption-related posts on SO. They observed that energy-related questions have distinct characteristics relative to the average SO questions. On average, they have 2.6 times more answers, are marked as favorites 3.89 times more often, have 68\% more visualizations, 10\% more ``up-votes'', and 11\% more comments. They identified seven major causes for energy consumption problems according to developers, and summarized eight common solutions suggested by developers to improve energy efficiency.
%Gantayat et al.~\cite{Gantayat2015} studied the synergy between voting and acceptance of answers on SO. They observed that in 81\% of threads that have multiple answers, the accepted answer is also the top-voted answer. The number of votes on a post may be influence askers' choice of acceptance, and vice versa. 
%Openja et al.~\cite{Openja2020} analyzed release engineering questions, to understand the modern release engineering topics of interest and their difficulty. 
 %Using topic modeling techniques, they found that (i) developers discussed on 38 release engineering topics covering all 6 phases of modern release engineering; (ii) the topics Merge Conflict, Branching \& Remote Upstream are more popular, while topics Code Review, Web Deployment, MobileApp Debugging \& Deployment, Continuous Deployment are less popular yet more complicated, (iii) Security is both popular and difficult.
Meng et al.~\cite{meng2018secure}, and Chen et al.~\cite{Meng2019:icse} examined SO posts on Java security, to reveal developers' concerns on security implementation, technical challenges, or vulnerabilities in answer code. %\textcolor{red}{An et al. ~\cite{An2017} investigated how SO might be used as a platform for code laundering, where code snippets are shared without proper validation, potentially leading to security vulnerabilities. }

Our study is related to all studies summarized above, but has a unique focus on the answer reuse by open-source Java projects available on GitHub. The study most related to our work was conducted by Nasehi et al.~\cite{Nasehi2012}, who manually inspected 163 discussion threads, and identified several characteristics of well-received answers (i.e., answers with score 4 or above). %They found that code examples and the accompanied explanation are two inseparable elements of well-received answers. 
%These characteristics include: making concise code examples, shaping the answer explanation based on questioners' expertise level, and making use of the question context to decrease the cognitive distance. 
%Although we did not manually inspect content of reused answers, 
Our data analysis complements Nasehi's study. 
%; it shows that answers with higher scores, larger code examples, and longer answer explanation are more likely to be reused.

\vspace{-.5em}
\subsection{Studies on SO--GitHub Connections} Studies were conducted to investigate the relationship between data mined from StackOverflow, and the data from GitHub~\cite{Badashian2014,An2017,Xiong2017,Yang2017,Baltes2017,Wu2019,Manes2019,Manes2021}. 
%\textcolor{red}{Manes and Baysal \cite{Manes2021} studied how code snippets from Stack Overflow evolve over time in GitHub projects, observing patterns in modifications and updates.}
Some studies associated users across platforms via common email addresses~\cite{Badashian2014,Xiong2017}. %explored links of developers' identities across two platforms 
With the associations,  
%Vasilescu et al.~\cite{Vasilescu2013} found that active GitHub committers asked fewer questions and provided more answers than others on SO. 
%active SO askers distributed their work in a less uniform way than developers that do not ask questions. 
Xiong et al.~\cite{Xiong2017} observed that active issue committers on GitHub are also active question askers. 
%(1) active issue committers on GitHub are also active question askers; (2) for most developers, the topics of their contents in GitHub are similar to those of their SO posts; (3) developers' concerns on SO shift as developers switch from one GitHub project to another; (4) developers' concerns on GitHub are more related to their answers than their questions or comments on SO. 
However, Badashian et al.~\cite{Badashian2014} showed that the relation between activities on the two platforms is not strong enough, to predict developers' activities on one platform based on their activities on the other. 
%Lee et al.~\cite{Lee2017} %analyzed the dataset of Badashian et al., and 
%to examine the similarities in developers' interests; they 
%reported that developers co-participating activities on the two platforms share common interests.
%  developers share similar interests with other developers who co-participated activities.
Some studies were performed to identify the reuse of SO answers by GitHub projects, via clone detection or keyword-based search~\cite{Yang2017,Baltes2017,Wu2019,Manes2019,Ragkhitwetsagul2021}. 

Our research also used clone detection to find answer reuse. However, 
different from the studies mentioned above, we did not blindly trust the clone detection results for two reasons. First, duplication 
does not necessarily imply code reuse. For instance, an old Java file can share code with a newly posted SO answer, although it is impossible for the file developers to take a time travel and refer to that answer posted in the future. 
Second, existing clone detectors can only find fragments that are very similar to each other. When code fragments are less similar, existing tools can fail to identify the reuse scenarios. To mitigate these issues of clone detection, we adopted two methods in our research. First, we manually inspected results of clone detection to remove false positives, which show code overlaps although developers actually did not reuse the answer code. Second, we also did keyword-based search to find Java files that explicitly reference the SO answers under analysis, 
%regular expression-based search to find Java files that explicitly reference SO answers, 
to identify some reuse scenarios missed by clone detection.

Manes and Baysal~\cite{Manes2019,Manes2021} mined SOTorrent and GHTorrent to locate files referencing SO posts.
%also used keyword-based search to locate GitHub projects that reference SO posts. %They showed that on average developers make 45 references to SO posts in their projects, with the highest number of references being made within the JavaScript code. 
They looked at 30 most popularly cited SO posts with non-programming language tags, and identified the top popular tags being related to Linux OS, string, and regular expressions~\cite{Manes2019}. They also analyzed the evolution patterns of reused code snippets on SO and GitHub~\cite{Manes2021} However, neither study  explores the similarity between referenced code on SO and the revised code on GitHub. 
Wu et al.~\cite{Wu2019} searched GitHub for source files with keyword ``stackoverflow'',
and manually inspected retrieved files written in Java, JavaScript, Objective C, PHP, or Python. 
%289 files from 182 open-source projects on GitHub that explicitly references SO posts. 
%By studying those files, 
They found that in 31.5\% of the data, developers modified source code from SO. In another 35.5\% of cases, developers used SO posts as an information source for later reference, instead of copying any code from those posts. 

As with Wu et al., we also observed different types of reuse practices by developers. However, the post distribution among reuse categories is different.
Namely, we found that developers modified code from SO in a much higher percentage of scenarios (i.e., 85\%); they referred to SO as an information source without copying any code in only 6\% of cases. 
This may be because our taxonomy of reuse patterns is more comprehensive and we formulated our dataset differently. 
Specifically, the taxonomy of Wu et al.~only includes 5 high-level categories: C1--C5; our taxonomy also includes 11 subcategories under C3, to further differentiate between developers' copy-paste-revise practices. In particular, among the 11 subcategories, we found C3.7 to be the largest one, for which developers copied and pasted most of the answer code while applying both statement-level updates and structure-level changes. 
Additionally, our dataset is larger, covering more unique files (i.e., 407 vs.~289) and repositories (i.e., 357 vs.~182); our dataset focused on Java code while Wu et al.~studied code in five languages.
%as we have more expertise in this language.

%However, we observed different distributions of the five reuse categories, probably because we studied a different dataset. Our taxonomy of reuse patterns is more comprehensive, because it defines more categories for the practices when answer code is modified before being used. 
%Different from Wu et al.'s study, 
Different from all prior work, our research characterizes the reused answers from new angles like question styles, post scores, post ages, code lengths, and text lengths; our paper introduces a new and carefully designed approach to precisely locate SO answers reused by Java files on GitHub.

\vspace{-.5em}
\section{Threats to Validity}

\paragraph{Threats to External Validity} Our study is based on the 30,000 SO answer posts retrieved by Google BigQuery, and the Java files from GitHub citing those answers or discussion threads.
%, as well as the Java files from GitHub that reference some of the retrieved answers or their discussion threads. 
Our findings may not generalize well to the threads or Java files not included in our dataset. To further investigate this threat,
we did keyword-based search on GitHub, and found around 108 thousand Java files citing SO threads as of 2024. Our dataset of 407 unique Java files citing SO threads is larger than 385, the minimum sample size required to derive representative conclusions  with a confidence level of 95\% for the large set of SO-citing Java files. Thus, our major findings are still representative and can generalize well.
%\textcolor{red}{While GitHub searches indicate a larger pool of around 108K Java files referencing OS threads as of 2024, statistical sampling methods were applied to select the dataset. The sample size strikes a balance between comprehensiveness and manageability. } 
%In the future, we plan to retrieve more answer posts with BigQuery, and even study developers' answer-reuse practices in non-Java programs and close-source software.

\vspace{-.5em}
\paragraph{Threats to Construct Validity}
We detected answer reuses in two complementary ways: clone detection and keyword-based search. 
%Clone detectors help identify answer reuses by revealing the similarity between code fragments; they can fail when developers reuse answers but do not produce code clones. Keyword-based search 
% help identify answer reuses by locating answer IDs explicitly mentioned in Java files; 
% they can fail when developers do not explicitly reference any SO answer. Therefore, 
There can be reuse scenarios not captured by either way, such as those having no clones and citing no answers explicitly. 
%where developers' code is very different from the reused code and the reused answer is not explicitly cited. 
The actual reuse scenarios existing in our dataset may be more than what we found. We share this limitation with existing studies. As more developers explicitly reference the SO answers they reuse in codebases, such limitations can get alleviated.
%expand our analysis scope to retrieve non-Java programs that 

\vspace{-.5em}
\paragraph{Threats to Internal Validity}
Our research involves manual inspection for (1) the refinement of clone-detection results and keyword-based search results, (2) topic identification for SO discussion threads, and (3) classification of developers' answer-reuse practices. 
%To mitigate the human error in this process, 
%However, 
Our manual analysis may be subject to human bias. To mitigate that issue, we had two authors (1) check the datasets simultaneously, and (2) discuss frequently to resolve any divergent opinions or labels.
%and does not scale to larger datasets. In the future, we plan to mitigate the threats by using more advanced clone detectors, and by creating tools that hardcode our insights to automate result refinement.

\vspace{-.5em}
\section{Conclusion}

We did an empirical study to characterize reused SO answers, and explore how Java developers reuse SO answers in GitHub projects. Compared with prior work, 
%our hybrid approach reveals answer reuse 
%the SO answers reused by Java files on GitHub 
%with more quality assurance. Our 
our hybrid approach is unique in two aspects. First, after using regular expressions to locate Java files that cite any SO post, it combines two methods---clone detection and keyword-based search---to detect candidate SO answers reused by Java files. 
As the two methods have separate strengths and weaknesses, combining them enables us to identify a high-quality set of answers potentially reused. Second, we did manual inspection to refine the candidates retrieved by both methods. 
%, to remove false positives where code fragments have accidental overlaps or Java files cite SO answers for non-reuse purposes. 
With such a careful approach design, we did our study with high rigor. In the future, we plan to create a tool to crawl for related code examples on GitHub given an SO answer.
%, to help developers better reuse answers.
%By showing the real-world examples, such tools can help developers better reuse answers. 

%\vspace{-1em}

\bibliographystyle{ACM-Reference-Format}
\bibliography{jun-workshop-2024}

\end{document}